\documentclass{optica-article}

\journal{opticajournal}

\articletype{Research Article}

\usepackage{lineno}
\usepackage{mathtools}
\usepackage{amsmath}
\usepackage{amsfonts}
\usepackage{caption}
\usepackage{subcaption}
\usepackage{blindtext}
\usepackage{algorithm}
\usepackage{algpseudocode}
\usepackage[capitalise]{cleveref}
\usepackage{booktabs}
\begin{document}

\title{A fast, large-scale optimal transport algorithm for holographic beam shaping}

\author{Andrii Torchylo,\authormark{1} Hunter Swan,\authormark{1} Lucas Tellez\authormark{2}, and Jason M. Hogan\authormark{1,*}}

\address{\authormark{1}Department of Physics, Stanford University, Stanford, California 94305, USA}
\address{\authormark{2}Department of Mathematics, Stanford University, Stanford, California 94305, USA}

\email{\authormark{*}hogan@stanford.edu}

\newtheorem{problem}{Problem}
\DeclarePairedDelimiter{\abs}{\lvert}{\rvert}
\DeclarePairedDelimiter{\norm}{\lVert}{\rVert}

\begin{abstract*} 
Optimal transport methods have recently established state of the art accuracy and efficiency for holographic laser beam shaping.  However, use of such methods is hindered by severe $\mathcal{O}(N^2)$ memory and $\mathcal{O}(N^2)$ time requirements for large scale input or output images with $N$ total pixels.  Here we leverage the dual formulation of the optimal transport problem and the separable structure of the cost to implement algorithms with greatly reduced $\mathcal{O}(N)$ memory and $\mathcal{O}(N\log N)$ to $\mathcal{O}(N^{3/2})$ time complexity.  These algorithms are parallelizable and can solve megapixel-scale beam shaping problems in tens of seconds on a CPU or seconds on a GPU. 
\end{abstract*}

\section{Introduction}
Laser beam shaping is a field with large and growing importance to a wide array of commercial and scientific applications, such as dipole trapping of neutral atoms \cite{brandt2011spatial}, quantum computing \cite{Endres6100}, and VR/AR technology \cite{blanche2021holography,SLMVRAR, NearEyeSLM}.  There are many techniques for shaping laser beams \cite{rosalesshape}, but holographic methods stand out for being essentially lossless while allowing arbitrary beam profiles \cite{Dickey,swan2025high}.  However, this comes at the cost of a challenging computational problem, known as phase retrieval: One must determine a phase to apply to a laser beam such that the desired target intensity profile is realized in the Fourier plane.

\begin{figure}[b]
    \centering
    \includegraphics[width=0.99\linewidth]{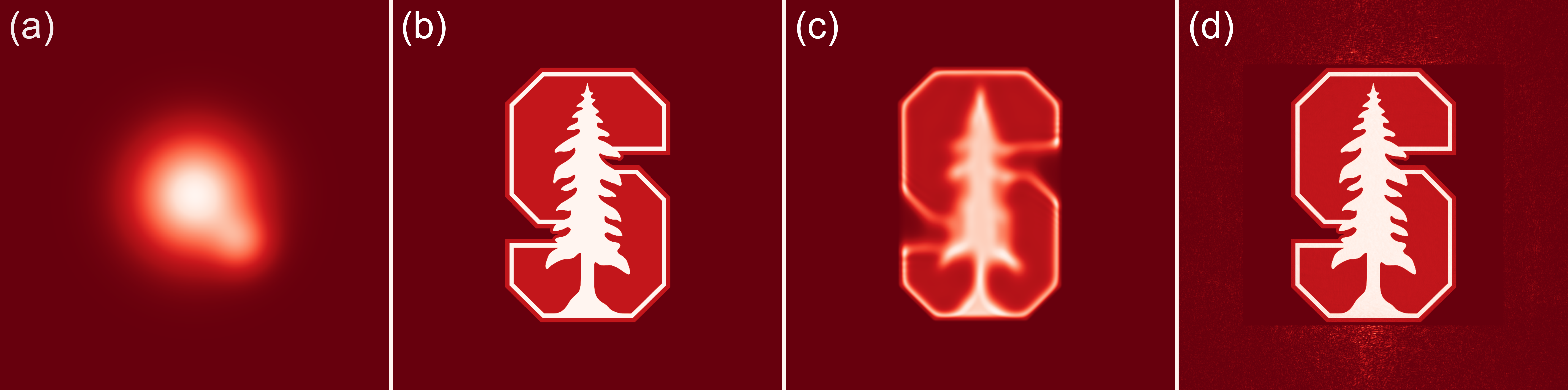}
    \caption{Simulated solution of an example beam shaping problem.  All images are 1024 by 1024 pixels.  \textbf{(a)} Input laser beam amplitude profile. The beam is a sum of two Gaussians, with deliberate asymmetry for illustration. \textbf{(b)} Target output beam profile. \textbf{(c)} Optimal transport solution produced by 200 iterations of the FOT algorithm introduced in this work. Solving required 76 seconds and 24 MiB of memory on an Intel Core i7 CPU, or 8.5 seconds on a Nvidia T4 GPU. \textbf{(d)} Solution produced by 100 MRAF iterations initialized from our FOT solution in (c).  Solving required a total time (FOT + MRAF) of 94 seconds on a CPU or 9.2 seconds on a GPU.}
    \label{fig1}
\end{figure}

There are many algorithms for solving this phase retrieval problem which make different tradeoffs \cite{gerchberg1972practical,Pasienski08,Harte14}.  Most significantly, some algorithms are able to boost the \textit{accuracy} with which a realized laser beam profile conforms to the desired target profile (as measured by e.g. $L^1$ or RMS intensity deviation) at the expense of reducing the \textit{efficiency} of the beam shaping process (as measured by the fraction of input light directed into the shaped output beam).  In previous work \cite{swan2025high}, we have demonstrated a method for generating an approximate solution to the phase retrieval problem which leads to final solutions that have state of the art accuracy and efficiency.  This method is based on optimal transport (OT), and can serve as a high quality initialization to conventional phase retrieval algorithms.  It has one major drawback, however, which is that it requires a prohibitive amount of memory and compute time for beam images larger than about 200 by 200 pixels.  This is due primarily to the need to store various matrices of size $N\times N$, with $N$ the total number of pixels, leading to an $\mathcal{O}(N^2)$ memory scaling. 

In this work, we exploit special structure of a reformulation of the underlying OT problem to produce an improved algorithm with $\mathcal{O}(N)$ memory scaling.  This resolves the main challenge of our previous algorithm, and has the additional benefit of making it much faster.  Our new algorithm can solve megapixel-scale holographic beam shaping problems in times of order 1 to 100 seconds on an Intel Core i7 CPU, depending on the complexity of the target profile.  Problems of such scale were impractical to solve without downsampling in our previous work. An example of the performance of this new algorithm is shown in Fig. \ref{fig1}.  Our algorithm is highly parallelizable, and we have demonstrated speedups of 10x on a Nvidia T4 GPU processor, which may allow for wide-ranging real time applications. 

\section{Overview in relation to previous work}
We recapitulate here the mathematical problem underlying laser beam shaping, as introduced in \cite{swan2025high}.  To distinguish our new algorithm from previous work, we will refer to the beam shaping algorithm of \cite{swan2025high}, which relied on a generic optimal transport library, as ``Black Box Optimal Transport'', or BBOT, and our new algorithm as ``Fast Optimal Transport'', or FOT.  We use lowercase letters to denote quantities associated with the plane of the input laser beam and uppercase letters for corresponding quantities in the Fourier plane.  Given are two arrays $g_{jk}$ and $G_{JK}$ representing the discretized amplitudes of the (measured) input laser beam and (desired) target output laser beam, respectively.  The values of these arrays are sampled on grids of points $x_j, y_k$ and $X_J, Y_K$, respectively, where $j,k,J,K \in\{0,\dots,n-1\}$ for some integer $n$.  We assume square arrays for simplicity, but this is not an essential limitation.  Without loss of generality, we may choose units such that the grid coordinate $x_j$ has the numerical value $x_j = (j-\lfloor n/2\rfloor )/\sqrt{n}$, and similarly for $y_k, X_J, Y_K$. 

We seek a discretized phase $\phi_{jk}$ such that 
\begin{equation}
\left|\mathcal{F}\left[g_{jk}e^{2\pi i\phi_{jk}}\right]_{LM}\right| \approx G_{LM},
\label{eq:main}
\end{equation}
where $\mathcal{F}$ is a discrete Fourier transform.  Note that $\phi_{jk}$ is a phase measured in cycles rather than radians.  The left-hand side of this relation represents the amplitude of a Fraunhofer diffraction pattern resulting from applying a phase $\phi$ on an input laser beam $g$.  The approximate equality asserts that this pattern should coincide with our target output beam.  More rigorously, we typically want to minimize the distance between the left- and right-hand side of \cref{eq:main} with respect to some preferred metric, such as $L^1$ or RMS distance. 

An approximate solution to \cref{eq:main} can be obtained by mapping it to an associated OT problem, in which we seek an array $\Gamma_{jkLM}$ (a transport plan) minimizing the total transport cost 
\begin{equation}
\sum_{jkLM} \Gamma_{jkLM} c_{jkLM},
\label{eq:total-cost}
\end{equation}
subject to a positivity constraint $\Gamma_{jkLM}\geq 0$ and marginal constraints $\sum_{jk} \Gamma_{jkLM} = G^2_{LM}$ and $\sum_{LM} \Gamma_{jkLM} = g^2_{jk}$. Here $c_{jkLM}$ is the discretized cost function, which has the form 
\begin{equation}
c_{jkLM} = \frac{\left(j-L\right)^2 + \left(k-M\right)^2}{2n^2}.
\label{eq:cost-function}
\end{equation}
Given such a $\Gamma_{jkLM}$, a solution $\phi_{jk}$ to the original problem \cref{eq:main} can be recovered via two summation operations.  The first such operation computes a discretized phase gradient $\left(\frac{\partial \phi}{\partial x}\right)_{jk}$, $\left(\frac{\partial \phi}{\partial y}\right)_{jk}$ from the first moments of $\Gamma$ via 
\begin{align}
\left(\frac{\partial \phi}{\partial x}\right)_{jk} & = \frac{1}{g^2_{jk}}\sum_{LM} \Gamma_{jkLM} X_L 
\label{eq:momentsx} \\
\left(\frac{\partial \phi}{\partial y}\right)_{jk} & = \frac{1}{g^2_{jk}}\sum_{LM} \Gamma_{jkLM} Y_M.
\label{eq:momentsy}
\end{align}
The second summation operation integrates these phase gradients via e.g. the trapezoid rule to compute the phase $\phi_{jk}$.  

In BBOT, the OT problem defined by \cref{eq:total-cost,eq:cost-function} is solved using a generic OT package \cite{OptimalTransportjl} as a ``black box''.  While convenient, such a generic OT solver is not able to leverage the special structure of our OT problem, and thus is severely performance limited for large grid sizes $n$.  In particular, storing the arrays $\Gamma_{jkLM}$ and $c_{jkLM}$ requires $n^4$ floating point numbers each, which for a 1 megapixel image and 64 bit precision becomes 8 terabytes of memory.

Our first improvement in the present work is to use the dual formulation of the above OT problem to eliminate the need to store the transport plan $\Gamma$.  Instead, we store two objects $u_{jk}$ and $V_{LM}$ of the same size as the input and output data, respectively, resolving the most challenging memory constraint of BBOT.  A second important insight is that the structure of the cost function in \cref{eq:cost-function} allows it to be decomposed as a sum of two lower dimensional arrays of the form $H_{jK} = (j-K)^2/2n^2$ (as noted e.g. in \cite{cuturiComputationalOT}), eliminating the need to ever store or compute the full cost array $c_{jkLM}$.  Expressing all operations for an OT solver in terms of the lower dimensional objects $u,V,H$ rather than $\Gamma, c$ results in dramatically fewer elementary operations and a corresponding improvement in time complexity.

\section{Overview of new algorithm}
In the entropic regularized dual formulation of an OT problem \cite{cuturi2013}, we seek optimization variables $u_{jk}$ and $V_{JK}$, which are related to the transport plan $\Gamma_{jkLM}$ via
\begin{equation}
\Gamma_{jkLM} = u_{jk} \exp\left(-c_{jkLM}/\varepsilon\right) V_{LM}
\label{eq:uv-to-plan}
\end{equation}
where the entropic regularization hyperparameter $\varepsilon$ has typical values $\varepsilon \sim 10^{-4}$ to $\varepsilon \sim 10^{-1}$.  Given the form of the cost function \cref{eq:cost-function}, the middle term in \cref{eq:uv-to-plan} can be written $\Lambda_{jL} \Lambda_{kM}$ with $\Lambda_{jK} = \exp\left(-H_{jK}/\varepsilon\right) = \exp\left(-\left(j-K\right)^2/2n^2\varepsilon\right)$.

The Sinkhorn-Knopp algorithm \cite{altschuler2017near} solves for $u_{jk}$ and $V_{JK}$ by repeatedly applying the update rules
\begin{align}
\label{eq:iter1}
u_{jk} & \leftarrow \frac{g^2_{jk}}{\sum_{LM}\Lambda_{jL} V_{LM} \Lambda_{kM}} = \Bigl[g^2 \oslash \left( \Lambda \cdot V \cdot \Lambda\right)\Bigr]_{jk} \\
\label{eq:iter2}
V_{JK} & \leftarrow \frac{G^2_{JK}}{\sum_{\ell m}\Lambda_{\ell J} u_{\ell m} \Lambda_{mK}} = \Bigl[G^2 \oslash \left( \Lambda \cdot u \cdot \Lambda \right)\Bigr]_{JK} \,\,,
\end{align}
where $\oslash$ denotes elementwise division and $\cdot$ denotes matrix multiplication.  These operations correspond to iteratively applying the two marginal constraints on the transport plan $\Gamma_{jkLM}$.  After applying these updates to convergence (discussed below), we use \cref{eq:momentsx,eq:momentsy} to compute the discretized gradient of the solution phase via
\begin{align}
    \label{eq:dphidx}
    &\left(\frac{\partial \phi}{\partial x}\right)_{jk} = \frac{u_{jk}}{g^2_{jk}}\sum_{LM} \Lambda_{jL} X_L V_{LM} \Lambda_{kM} =  \Bigl[u \odot \left(\Lambda \cdot \text{diag}(X) \cdot V \cdot \Lambda \right)\oslash g^2\Bigr]_{jk} \\
    \label{eq:dphidy}
    &\left(\frac{\partial \phi}{\partial y}\right)_{jk} = \frac{u_{jk}}{g^2_{jk}}\sum_{LM} \Lambda_{jL} V_{LM} Y_M \Lambda_{kM} = \Bigl[u \odot \left(\Lambda \cdot V \cdot \text{diag}(Y) \cdot \Lambda  \right)\oslash g^2\Bigr]_{jk} \,\,,
\end{align}
where $\odot$ denotes elementwise multiplication.  Note that here we have eliminated the transport plan $\Gamma_{jkLM}$ using \cref{eq:uv-to-plan}, avoiding the need to explicitly store this large array in memory. 

The remaining integration step to compute $\phi_{jk}$ from $\frac{\partial \phi}{\partial x}$ and $\frac{\partial \phi}{\partial y}$ is identical to \cite{swan2025high}.  Our full phase retrieval algorithm is summarized in \cref{alg:fast-2d-sinkhorn}. 

\begin{algorithm}[!t]
    \caption{Fast Optimal Transport (FOT)}
    \label{alg:fast-2d-sinkhorn}
    \begin{algorithmic}[1]
    \State $u \gets \mathbf{1}_{n\times n}/n^2$
    \State $V \gets \mathbf{1}_{n\times n}/n^2$
    \While {not converged}
        \State $u \gets g^2 \oslash (\Lambda \cdot V \cdot \Lambda)$ \Comment{Equation \ref{eq:iter1}} 
        \State $V \gets G^2 \oslash (\Lambda \cdot u \cdot \Lambda)$ \Comment{Equation \ref{eq:iter2}} 
    \EndWhile

    \State $\partial\phi/\partial x \gets u \odot (\Lambda \cdot \operatorname{diag}(X) \cdot V \cdot \Lambda) \oslash g^2$ \Comment{Equation \ref{eq:dphidx}} 
    \State $\partial\phi/\partial y \gets u \odot (\Lambda \cdot V \cdot \operatorname{diag}(Y)\cdot \Lambda) \oslash g^2$ \Comment{Equation \ref{eq:dphidy}} 
 
    \State $\phi = \text{Integrate}(\partial\phi/\partial x, \partial\phi/\partial y)$ \Comment{Described in \cite{swan2025high}} 
    \State \Return $\phi$ 
    \end{algorithmic}
\end{algorithm}

The convergence criterion for \cref{eq:iter1,eq:iter2} can be taken to be any of several loss functions.  In particular, total variation of marginal constraint violation has theoretical guarantees of convergence \cite{altschuler2017near} and can be computed quickly (see Supplementary Material). In practice, one often simply runs a fixed number of iterations without computing any loss function. In the Supplementary Material we provide loss function definitions and various further practical details on implementation, along with a self-contained Julia version of \cref{alg:fast-2d-sinkhorn}. 

\begin{figure}[b]
    \centering
    \includegraphics[width=0.49\linewidth]{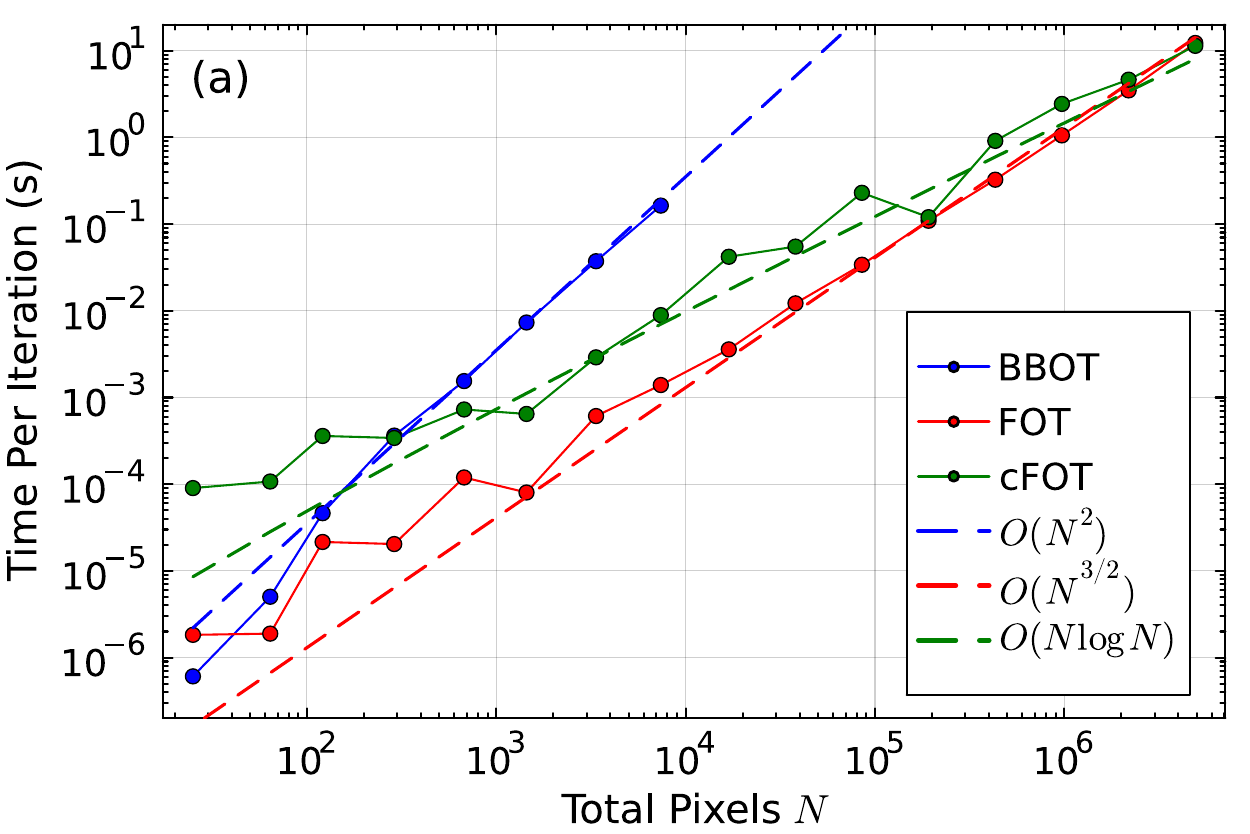}
    \includegraphics[width=0.49\linewidth]{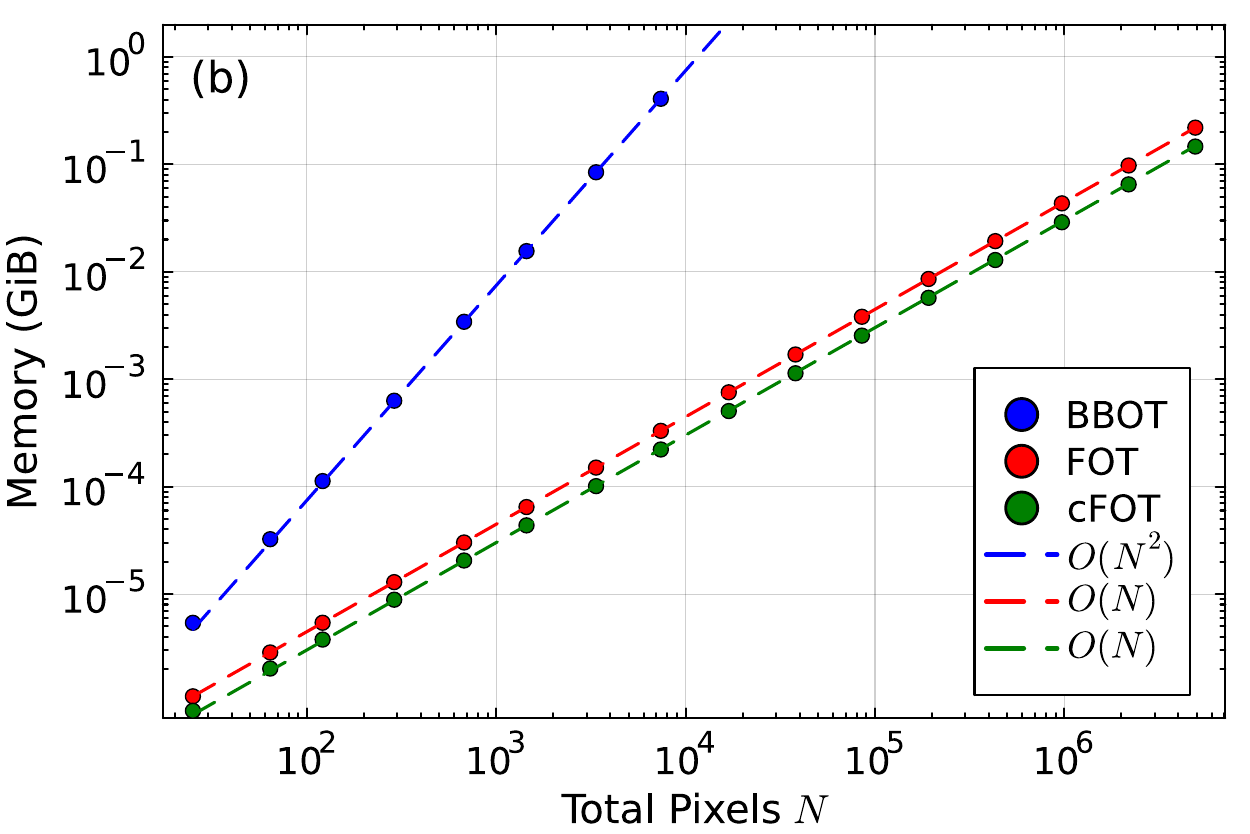}
    \caption{Scaling of \textbf{(a)} time per iteration and \textbf{(b)} memory of BBOT, FOT, and cFOT algorithms.  Dashed lines show various asymptotic scalings as a guide to the eye.  Input and target distributions have identical shapes to those of \cref{fig1}. Time per iteration is computed with \texttt{btime} from \cite{BenchmarkTools}. Memory usage is computed using Julia's built-in \texttt{sizeof} function for all stored variables in the inner loop of \cref{alg:fast-2d-sinkhorn}.}
    \label{fig:scaling}
\end{figure}

\subsection{Convolutional speedup}
Due to the fact that $\Lambda_{jK}$ is a Toeplitz matrix (i.e. $\Lambda_{jK}$ depends only on $j-K$), for sufficiently large image sizes $n$, the matrix multiplication operations in \cref{eq:iter1,eq:iter2} can be evaluated more quickly with a 1D linear convolution.  In terms of the vector $\lambda_j = \exp\left(-j^2/2n^2\varepsilon\right)$ with indices $j=-(n-1),-(n-2),\dots,(n-1)$, one can show 
\begin{equation}
\Lambda \cdot V \cdot \Lambda = \lambda * V * \lambda,
\end{equation}
where $\lambda * V$ or $V * \lambda$ denotes a linear convolution of $\lambda$ with columns or rows of $V$ respectively (see Supplementary Material). 
The same approach can clearly be applied to \cref{eq:dphidx,eq:dphidy} as well.  We refer to the variant of \cref{alg:fast-2d-sinkhorn} using linear convolutions in place of matrix multiplication as ``Convolution Fast Optimal Transport'', or cFOT, to disambiguate it from FOT. 

\section{Complexity Analysis}

FOT only stores three arrays $u$, $V$, $\Lambda$, each having size equal to that of the input and output images, i.e. $n\times n =: N$.  Therefore the memory complexity is $\mathcal{O}(N)$.  cFOT requires slightly less memory by storing $\lambda$ rather than $\Lambda$. 

Elementwise operations $\odot$ and $\oslash$ require $n^2$ basic operations, while matrix multiplication and 1D linear convolution require $\mathcal{O}(n^3)$ and $\mathcal{O}(n^2 \log(n))$ basic operations, respectively. Thus, Sinkhorn-Knopp iterations \cref{eq:iter1,eq:iter2} are $\mathcal{O}(n^3)$, and the convolutional variant is $\mathcal{O}(n^2\log n)$. By comparison, each iteration of the generic OT solver of BBOT involves a vector-matrix product of the full $n^2\times n^2$ cost matrix, which requires $\mathcal{O}(n^4)$ basic operations \cite{OptimalTransportjl, POT}.  Similarly, the phase gradient calculation \cref{eq:dphidx,eq:dphidy} is now $\mathcal{O}(n^3)$ for FOT or $\mathcal{O}(n^2\log n)$ for cFOT vs. $\mathcal{O}(n^4)$ for BBOT. In all cases the final integration step is $\mathcal{O}(n^2)$.  

For a fixed number of iterations, the total time complexity of the new algorithm is $\mathcal{O}(n^3)=\mathcal{O}(N^{3/2})$ for FOT and $\mathcal{O}(n^2 \log n) = \mathcal{O}(N \log N)$ for cFOT.  A theoretical analysis \cite{altschuler2017near} has provided a bound of $\mathcal{O}(\delta^{-2} \log N)$ iterations for convergence to a tolerance $\delta$, as measured by the total variation of marginal constraint violation in the OT problem (see Supplementary Material). Our analysis implies an overall time complexity of $\mathcal{O}(\delta^{-2}N^{3/2}\log N)$ or $\mathcal{O}(\delta^{-2}N\log^2 N)$ to fixed convergence for the two variants.  This is to be compared to $\mathcal{O}(N^2)$ per iteration and $\mathcal{O}(\delta^{-2}N^2\log N)$ to fixed convergence for BBOT.  Memory and time complexity for BBOT, FOT, and cFOT is summarized in \cref{tab:complexity_comparison}.

\begin{table}[t]
    \centering
    \small
    \renewcommand{\arraystretch}{1.05} 
    \caption{Asymptotic computational resource requirements of the OT algorithms considered in this work. Here $N = n^{2}$ is the total number of pixels and $\delta$ is convergence tolerance.}
    \label{tab:complexity_comparison}
    \vspace{0.25em}
    \begin{tabular}{lccc}
        \toprule
        \textbf{Method} & \textbf{Memory} & \textbf{Time per iteration} & \textbf{Time to convergence} \\
        \midrule
        BBOT & $\mathcal{O}(N^{2})$ & $\mathcal{O}(N^{2})$ & $\mathcal{O}(\delta^{-2}N^2\log N)$ \\
        FOT  & $\mathcal{O}(N)$     & $\mathcal{O}(N^{3/2})$ & $\mathcal{O}(\delta^{-2}N^{3/2}\log N)$ \\
        cFOT & $\mathcal{O}(N)$     & $\mathcal{O}(N \log N)$ & $\mathcal{O}(\delta^{-2}N\log^2 N)$ \\
        \bottomrule
    \end{tabular}   
\end{table}

\begin{figure}[t]
    \centering
    \includegraphics[width=0.99\linewidth]{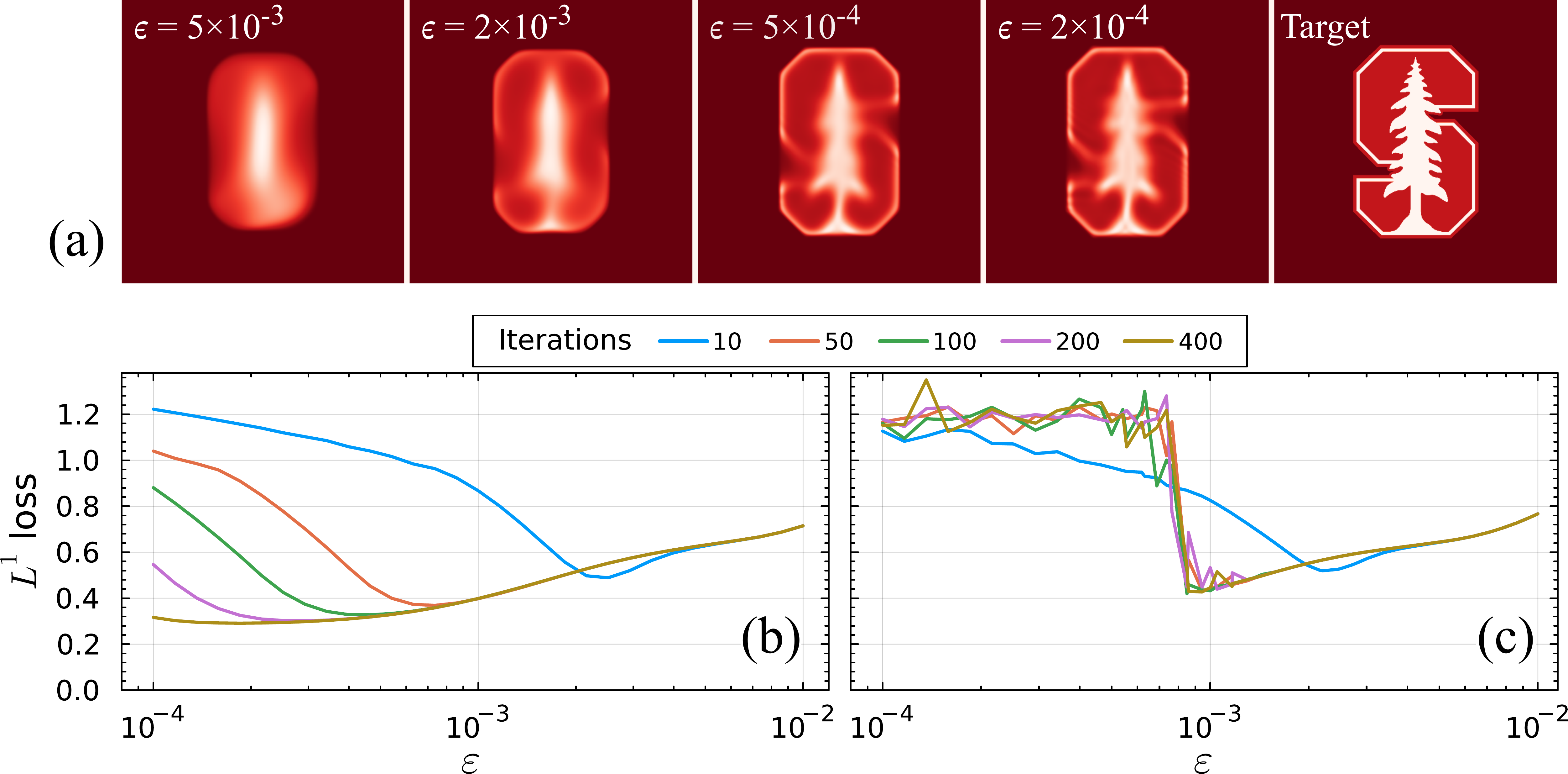}
    \caption{\textbf{(a)} Output beams from 200 iterations of FOT at different values of $\varepsilon$, with no subsequent polishing by other algorithms such as GS, MRAF, or CFM.  The input and target beams are those of \cref{fig1} at a resolution of $512\times 512$ pixels. \textbf{(b)} $L^1$ distance between target intensity and output beam intensity generated by FOT vs. $\varepsilon$ for different numbers of iterations.  Input and target beams are the same as in (a).  \textbf{(c)} Same as (b) but for cFOT.}
    \label{fig:epsilon}
\end{figure}

\section{Performance}

\cref{fig:scaling} shows scaling of memory and time per iteration versus the total number of pixels of the input and target images, demonstrating the expected $\mathcal{O}(N^{3/2})$ and $\mathcal{O}(N \log N)$ scaling respectively for the FOT and cFOT algorithms, along with the steeper $\mathcal{O}(N^2)$ scaling of BBOT. 

Our algorithm has a single hyperparameter $\varepsilon$, which controls the amount of entropic regularization to the OT loss landscape \cite{cuturi2013}.  As shown in \cref{fig:epsilon}, larger values can provide stronger and faster convergence but have a larger final error.  \cref{fig:epsilon}(b) and \cref{fig:epsilon}(c) show the $L^1$ intensity loss (i.e. the $L^1$ norm of the difference $\left|\mathcal{F}\left[g_{jk}e^{2\pi i \phi_{jk}}\right]\right|^2_{JK} - G^2_{JK}$) as a function of $\varepsilon$ for FOT and cFOT phase solutions, respectively.  Setting $\varepsilon$ too low can lead to numerical instability in the Sinkhorn-Knopp iterations \cref{eq:iter1,eq:iter2}, as seen in \cref{fig:epsilon}(c) for cFOT solutions near $\varepsilon \sim 10^{-3}$.  We thus typically set $\varepsilon$ to be as low as possible while avoiding such instability.  In our FOT implementation, this threshold is typically around $\varepsilon\sim 0.0002$. 

Fully saturating the algorithm's convergence is often not necessary, as our algorithm is typically only the first step of a phase retrieval procedure which also involves a polishing step using a conventional phase retrieval algorithm like Gerchberg-Saxton (GS) \cite{hirsch1971method,gerchberg1972practical}, mixed region amplitude freedom (MRAF) \cite{Pasienski08}, or cost function minimization (CFM) \cite{Harte14}.  As shown in \cite{swan2025high}, initializing these polishing algorithms with an OT solution leads to better accuracy and efficiency of the final phase solution. The accuracy required of an OT solution must only be sufficient that subsequent polishing iterations do not spawn phase vortices.

\begin{figure}
    \centering
    \includegraphics[width=0.99\linewidth]{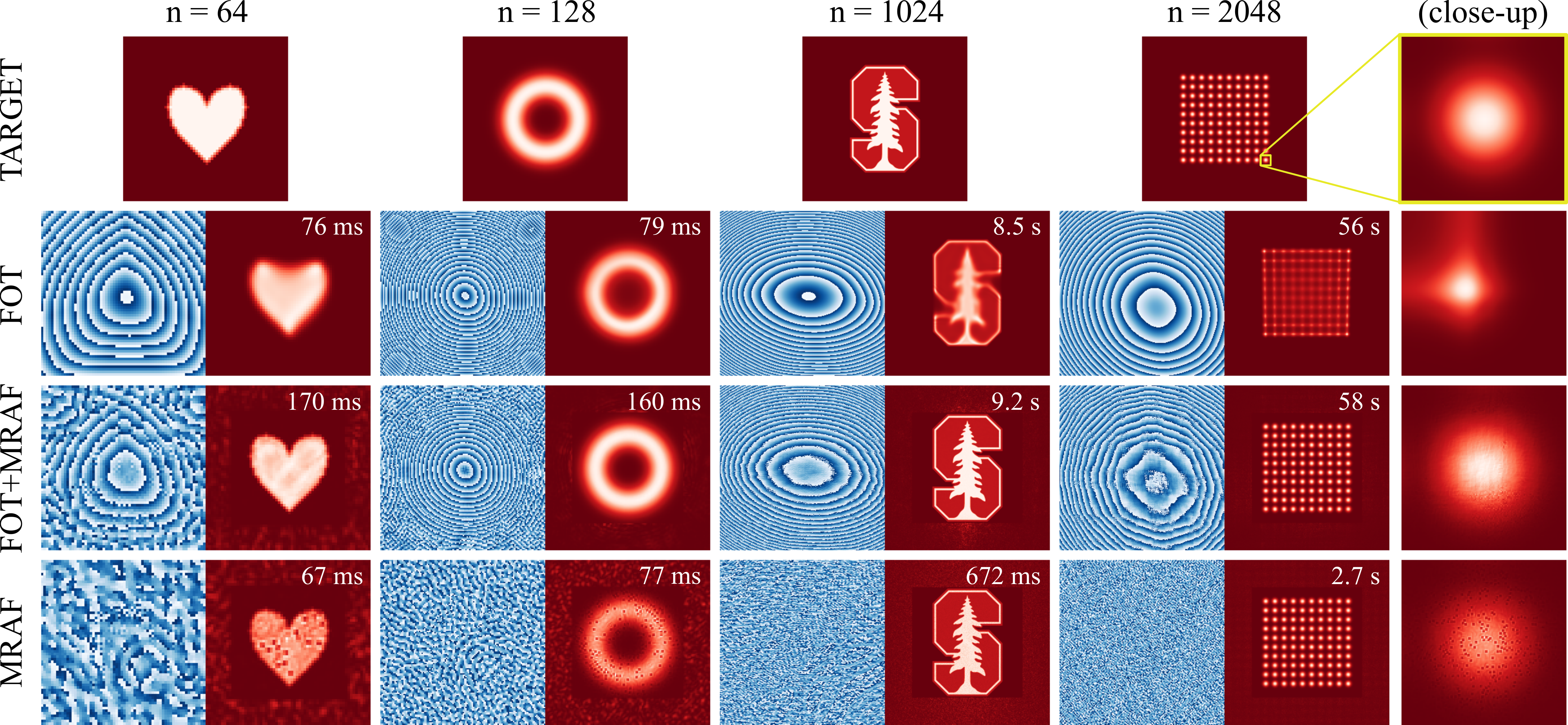}
    \caption{Phase solutions (blue) and corresponding output beams (red) from FOT, FOT polished with MRAF, and MRAF initialized with a flat phase. For each output beam, the inset time is the total computation time for generating the phase solution on an Nvidia T4 GPU. The input beam is that shown in \cref{fig1}, resized appropriately. FOT was run for 200 iterations with $\varepsilon=0.0002$ and MRAF was run for 100 iterations.  For the $n=1024$ and $n=2048$ phases, only the central $256\times 256$ region is shown.  For the highest resolution target, the rightmost column shows a close-up view at full resolution of a $128 \times 128$ pixel region surrounding the bottom right spot of the array.  Phase vortices in MRAF solutions can be identified by diffraction-limited dark spots in the output beam or corresponding points with non-zero winding number on the phase solution.}
    \label{fig:targets}
\end{figure}

FOT is capable of working directly with target images of much higher resolution than was previously possible with BBOT.  Phase solutions for a variety of targets of size $n=64$ ($N=4096$) up to $n=2048$ ($N=4194304$) are shown in \cref{fig:targets}, demonstrating that the phases remain vortex free in regions of appreciable light intensity after MRAF polishing even at the highest resolutions.  Phase solutions generated by MRAF initialized with a uniform phase are also shown in \cref{fig:targets} for comparison, exhibiting the phase vortices typical of other phase retrieval methods. 

The Sinkhorn algorithm is intrinsically parallelizable\cite{cuturi2013}, as are all additional operations in our algorithm.  We have demonstrated a 10-fold speedup by deploying our algorithm on a Nvidia T4 GPU.  As shown in \cref{fig:targets}, FOT can solve a $128\times 128$ pixel phase retrieval problem with 200 iterations in under 100~ms.  Such speeds may allow for real time application in many settings. 

\section{Conclusions}

We have introduced a new optimal transport-based algorithm for solving phase retrieval problems for holographic beam shaping which relaxes the memory and time complexity constraints of previous work. As with other OT methods, this algorithm is best used in combination with an iterative Fourier transform algorithm or cost function minimization algorithm for subsequent polishing to high solution accuracy.  Our new method allows for the solution of realistic megapixel-scale phase retrieval problems in timescales of a few seconds and peak memory allocation of tens of megabytes.  This performance is sufficient for real time application in some settings, such as dipole trapping of neutral atoms.  Another benefit of our method is that it is self-contained, not relying on any black-box OT solvers, which may facilitate its adoption in settings where existing OT packages are not available. 

\begin{backmatter}
\bmsection{Funding}
This work was supported by the Gordon and Betty Moore Foundation Grant GBMF7945 and the NSF QLCI Award No. OMA-2016244.
\bmsection{Disclosures}
The authors declare no conflicts of interest.

\bmsection{Data availability} 
No data were generated or analyzed in the presented research. Code for generating all figures and graphs are available upon reasonable request.
\bmsection{Supplemental document}
See Supplement 1 for supporting content.

\end{backmatter}

\bibliography{main.bib}

\end{document}